\documentclass[aps,prl,twocolumn, showpacs,preprintnumbers,amsmath,amssymb,oneside,floatfix]{revtex4-1}

\usepackage{graphicx}
\usepackage{dcolumn}
\usepackage{bm}

\begin{document}


\title{Self-phase modulation of a single-cycle terahertz pulse by nonlinear free-carrier response in a semiconductor}

\author{Dmitry Turchinovich$^{1,2}$, J{\o}rn M. Hvam$^1$, and Matthias C. Hoffmann$^3$}

\email{dmtu@fotonik.dtu.dk}

\affiliation{ $^1$DTU Fotonik, Technical University of Denmark, DK-2800 Kgs.~Lyngby,
Denmark\\
$^2$Max Planck Institute for Polymer Research, Ackermannweg 10, 55128 Mainz, Germany\\
$^3$SLAC Linear Accelerator Laboratory, 2575 Sand Hill Road, Menlo Park, CA, 94025, U.S.A.
}%


\begin{abstract}
We demonstrate the self-phase modulation (SPM) of a single-cycle THz pulse in a semiconductor, using bulk \emph{n}-GaAs as a model system. The SPM arises from the heating of free electrons in the electric field of the THz pulse, leading to an ultrafast reduction of the plasma frequency, and hence to a strong modification of the THz-range dielectric function of the material. THz SPM is observed directly in the time domain. In the frequency domain it corresponds to a strong frequency-dependent refractive index nonlinearity of \emph{n}-GaAs, found to be both positive and negative within the broad THz pulse spectrum, with the zero-crossing point defined by the electron momentum relaxation rate. We also observed the nonlinear spectral broadening and compression of the THz pulse.

\end{abstract}


\maketitle

Since its first demonstration \cite{1st_NLO} more than five decades ago, nonlinear optics (NLO) has revolutionized the field of photonics.
Until recently, ultrafast NLO was largely confined to the visible and infrared spectral ranges. However, with the emergence of sources generating high-energy ultrafast single-cycle terahertz (THz) pulses \cite{Matze_LNB, hebling_SPM, hoffmann_fulop}, more and more NLO studies are being conducted in the THz range. Nonlinear dynamics in semiconductors in strong THz fields is a particularly rich area of research due to good coupling between the THz fields and free carriers, and due to the possibility of inducing strong polarization effects in semiconductors with strong THz signals \cite{Ganichev}. Many interesting ultrafast phenomena have been investigated recently, such as interaction of strong THz fields with excitons \cite{Koch, tanaka_excitons}, THz-induced impact ionization of electrons in narrow-gap semiconductors \cite{Lindenberg, matze_impact}, ultrafast THz-induced electron dynamics in complex conduction bands \cite{Matze_transp1, razzari2009, Matze_transp2, koch_hegmann}, and femtosecond all-optical switching in nano-scale structures with THz signals \cite{dmt_mch_THz_QCSE, shimano_CNT}.

The majority of studies in the present-day ultrafast THz NLO are focused on achieving a better understanding and control of nonlinear processes in (mostly semiconductor) materials using THz light. There are, however, very few reports aiming at control and manipulation of THz light making use of the material's nonlinear response. Yet, the ultrafast THz signals are usually single-cycle waveforms, which can be experimentally sampled with a time resolution much better than one optical cycle, providing the complete time-domain information about the optical signal - its amplitude \emph{and} phase \cite{tonouchi, jepsen_LPR}. As we will show in this Letter, this makes the THz pulses ideal tools for the direct observation of both \emph{i)} the THz-specific NLO phenomena, and \emph{ii)} the general NLO phenomena occurring in the single-cycle optical regime.

Single-cycle pulses can not be described by the traditional carrier-envelope approximation \cite{Krausz_single_cycle}. Irrespective of the frequency range to which they belong, they inherently have an extremely broad spectral bandwidth covering many octaves of frequencies. Hence, the propagation of such signals in the medium is governed by the complex-valued material dielectric function $\hat{\varepsilon}(\omega)$ covering a very broad spectral range.
An optical nonlinearity - the near-instantaneous modification
of this dielectric function due to a nonlinear light-matter
interaction occurring in the single-cycle regime, corresponds to a
modification of the optical properties of matter over an extremely
broad range of frequencies. We find the effect of such a sudden change of dielectric environment on the propagating single-cycle waveform to be quite intriguing, which is the motivation behind the present work.

In this Letter we report on the study of a fundamental nonlinear optical effect - the self phase modulation (SPM) \cite{SPM1, SPM2}, of a single-cycle THz pulse. The SPM occurs due to the change of the (real part of) refractive index $n(\omega)= \text{Re} \; \sqrt{\hat{\varepsilon}(\omega)}$ of the material as a result of light-matter interaction. As an efficient nonlinear medium for the THz range we used an \emph{n}-doped bulk semiconductor GaAs, whose optical nonlinearity arises from the response of free electrons to the THz fields. The \emph{n}-GaAs was chosen because it is a well known, model system; however, any other doped or photoexcited semiconductor with a complex band structure will also exhibit the THz-range SPM according to the mechanism described in this Letter. The SPM was observed by us directly in the time domain using THz time-domain spectroscopy (TDS) \cite{tonouchi, jepsen_LPR}, and was analyzed in the frequency domain. In particular, we found that the nonlinear refractive index $\Delta n (\omega)$ of \emph{n}-GaAs is strongly frequency-dependent: it exhibits both positive \emph{and} negative nonlinearity within the broad spectral bandwidth of a single-cycle THz pulse, with the sign of nonlinearity determined by the the rate of electron momentum relaxation in the semiconductor. We also observed the nonlinear spectral broadening and compression of the THz pulses in the material.

\begin{figure} [t]
\centering
   \includegraphics[width=6.5cm]{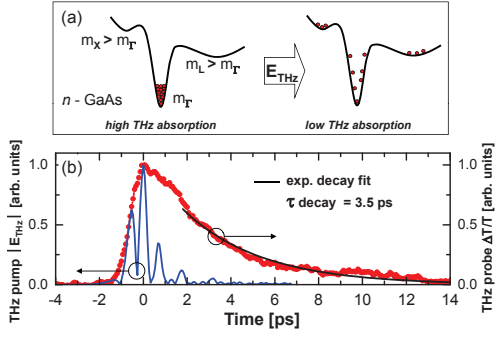}
  \caption
  {\label{fig1} (a) Mechanism of THz nonlinearity in \emph{n}-GaAs, based on carrier heating in the electric field of a THz pulse, leading to a decrease of plasma frequency $\omega_p$, and saturation of THz absorption. (b) Temporal dynamics of THz nonlinearity in \emph{n}-GaAs, observed in a transmission-mode TPTP measurement. Near-instantaneous bleaching of the sample via carrier-heating in the THz field is followed by a slower recovery due to electron cooling.}
   \end{figure}

In this Letter we show that the THz-range optical nonlinearity of a doped or photoexcited semiconductor is based on the modification of its plasma frequency by the carrier heating in the THz field. Semiconductors with free carriers are good absorbers of THz radiation, and their THz-range complex-valued dielectric function $\hat \varepsilon (\omega)$ is well described by the basic Drude plasma model \cite{Dressel, huber_nature}

\noindent \begin{equation}\label{eq1} \hat \varepsilon (\omega) =
(n + i\alpha c / 2 \omega)^2 = \varepsilon_{dc} - \omega_p^2
/(\omega^2 - i \omega / \tau_{r})
\end{equation}

\noindent or by one of its extensions \cite{Smith, Nemec}. Here $n$ and $\alpha$ are the frequency-dependent refractive index and power
absorption coefficient, respectively, $\tau_{r}$ is a carrier
momentum relaxation time, and $\varepsilon_{dc}$ is the background
"static" dielectric constant of the semiconductor in the absence of
free carriers \cite{justification_epsilon_dc}. $\omega_p = (N e^2 / \epsilon_0 m)^{1/2}$ is the plasma frequency, where $e$ is the elementary charge, $N$ is the free carrier density, $\epsilon_0$ is the vacuum permittivity, and $m$ is the carrier effective mass.

It directly follows from Eq. (\ref{eq1}) that the change in plasma frequency $\omega_p$ resulting from the THz light -- matter interaction will lead to a change in the THz absorption $\alpha$ and the refractive index $n$ of the semiconductor, i.e. to nonlinear absorption and SPM.  In the limiting case of $\omega_p \rightarrow 0$ the THz absorption will vanish, and the THz refractive index will be defined by the background dielectric constant $n(\omega_p \rightarrow 0) = \sqrt{\varepsilon_{dc}}$. The only mechanism to reduce the plasma frequency in a doped semiconductor (i.e. where $N$ is fixed), is by increasing the carrier effective mass $m$. In \emph{n-}doped materials (e.g. \emph{n}-GaAs) this can be achieved by electron heating in a sufficiently strong electric field - by redistribution of the electron population in energy-momentum space of the conduction band from low-momentum, small-effective-mass states around the bottom of the $\Gamma$-valley to high-momentum, large-effective mass states in the satellite valleys and to the states of strong nonparabolicity within the $\Gamma$-valley  (see e.g. \cite{Brennan_Hess} and references therein). In \emph{p}-doped materials the hole scattering between the hole subbands can be utilized.

\begin{figure} [t]
\centering
   \includegraphics[width=7.5cm]{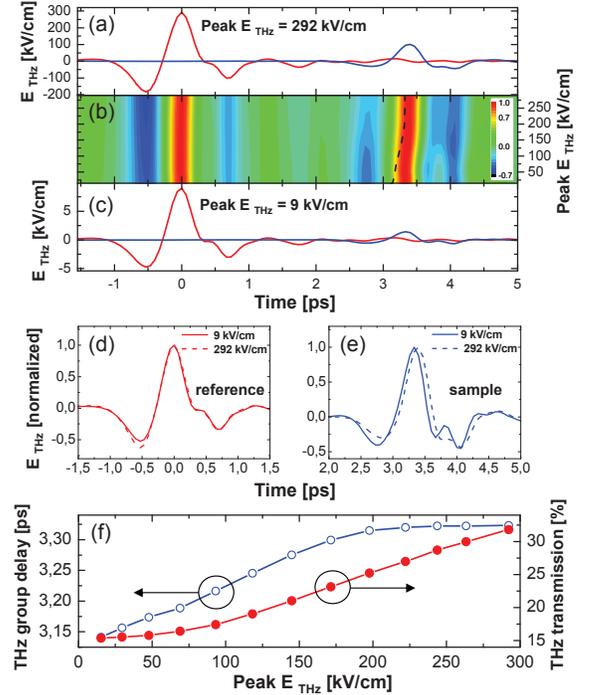}
  \caption
  {\label{fig2} THz SPM in \emph{n}-GaAs in the time domain. Peak $E_{THz}$ of reference THz pulse in (a) is 292 kV/cm, and in
  (c) is 9 kV/cm. (b) Reference and sample pulses normalized to their maxima, for the whole range of peak $E_{THz}$ of 9 - 292 kV/cm. Black dashed line: estimated group delay of the sample pulse. (d,e) Normalized reference (d) and sample (e) pulses for the cases of peak $E_{THz}$ of 9 kV/cm and 292 kV/cm, demonstrating the SPM in the sample. (f) Amplitude frequency-integrated transmission coefficient, and estimated group delay of a reference pulse as a function of peak $E_{THz}$. }
   \end{figure}

Recently the carrier heating in the ponderomotive potential created by the electric field of a single-cycle THz pulse \cite{Matze_transp1, razzari2009, Matze_transp2, koch_hegmann} was demonstrated. In \cite{Matze_transp1} it was shown that right after the excitation of \emph{n}-GaAs by a THz pulse with peak electric field $E_{THz} \simeq$ 100 kV/cm, the effective mass of electrons initially occupying the bottom of $\Gamma$-valley  ($m_{\Gamma} = 0.063 m_0$) drastically increases and reaches the values of $m \simeq 0.3 m_0$, more typical for the $L-$ and $X-$valley states in GaAs ($m_L = 0.23 m_0$, and $m_X = 0.43 m_0$) [$m_0$ is the electron rest mass] \cite{Brennan_Hess, ponderomotive}. The THz saturable absorption based on this mechanism, shown in Fig. \ref{fig1}(a), was observed by us in various \emph{n}-doped semiconductors \cite{dmt_mch_THz_satabs}. The temporal response of THz nonlinearity of \emph{n}-GaAs is near-instantaneous, as shown in Fig. \ref{fig1}(b). This measurement is a result of THz pump - THz probe (TPTP) spectroscopy \cite{Matze_transp1, Matze_transp2, koch_hegmann} where our \emph{n}-GaAs sample with the thickness $d$ = 0.4 mm, $N = 5 \times 10^{15}$ cm$^{-3}$, and $\tau_{r}$ = 94 fs was excited by a THz pump pulse with peak $E_{THz} \simeq 200$ kV/cm.
The near-instantaneous bleaching of the sample is followed by a slower recovery as the electrons return to the bottom of the $\Gamma$-valley \cite{Planken}. This, and all other measurements in this work were performed at room temperature.

We observed the SPM of single-cycle THz pulses in our \emph{n}-GaAs
sample in a nonlinear THz TDS experiment set up in
transmission configuration \cite{dmt_mch_THz_satabs}. The frequency spectrum of the THz pulses, generated by optimized optical rectification of 800 nm, 100 fs lasers pulses in lithium niobate (LN) \cite{Matze_LNB, hebling_SPM, hoffmann_fulop}, covered the
range 0.1 - 3 THz, thus containing many octaves of frequencies. The
peak electric field $E_{THz}$ in the generated THz pulses was
adjusted in the range of 9 - 292 kV/cm using a pair of wire-grid
polarizers positioned after the THz emitter in the spectrometer
\cite{dmt_mch_THz_satabs, Matze_LNB}. THz transmission through the polarizers is nominally frequency-independent. The THz pulses were recorded in the time domain using free-space electro-optic sampling \cite{tonouchi, jepsen_LPR}.

\begin{figure} [t]
\centering
   \includegraphics[width=6.5cm]{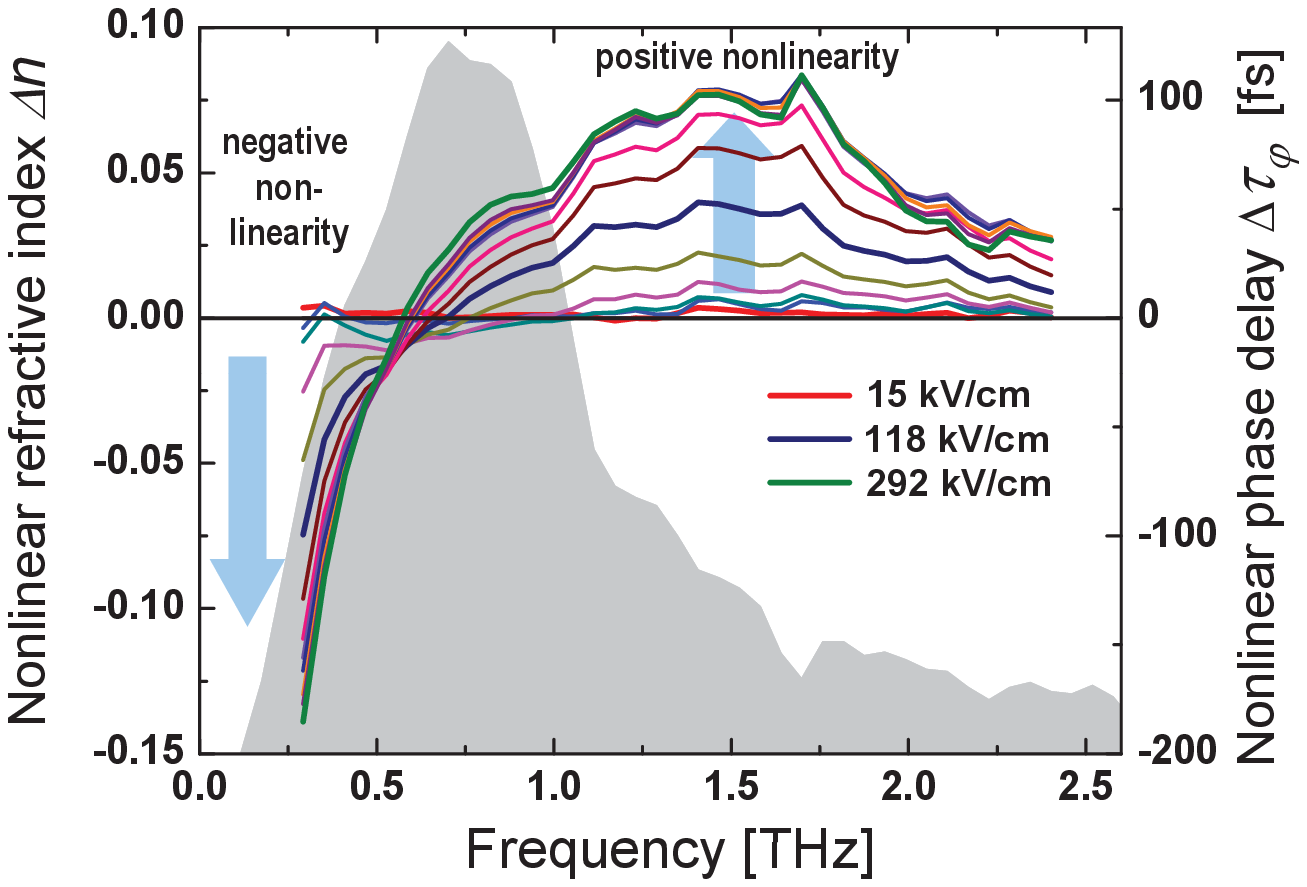}
  \caption
  {\label{fig3} THz SPM in \emph{n}-GaAs in spectral domain. Nonlinear contribution to refractive index $\Delta n$, and corresponding nonlinear phase delay as a function of frequency and peak  $E_{THz}$ in the range 9 - 292 kV/cm. Blue arrows show the positive and negative refractive index nonlinearity with increasing peak $E_{THz}$. The amplitude spectrum of reference THz pulse is shown as a background.}
  \end{figure}

As shown in Fig. \ref{fig2}, for each value of peak $E_{THz}$ in the generated THz pulse a reference (travelling through the optical path in the spectrometer in vacuum) and a sample (travelling through the same geometrical path, but with the sample inserted into it) THz pulses \cite{tonouchi, jepsen_LPR}, centered around the time delays of 0 ps and 3.3 ps, respectively, were recorded. A clear THz saturable absorption (i.e. increase in THz transmission)\cite{dmt_mch_THz_satabs} in our \emph{n}-GaAs sample with growth in peak electric field $E_{THz}$ is observed. In Figs. \ref{fig2}(a) and (c) the measured reference and sample THz pulses for the two extreme values of peak $E_{THz}$, 292 kV/cm and 9 kV/cm, are shown as examples of raw data. Here and below all mentioned peak values of $E_{THz}$ correspond to that of the reference THz pulse, and are reported as measured at the position of the sample.

In Fig. \ref{fig2}(b) a two-dimensional plot shows the reference
and sample pulses normalized to their respective peak values, for
the whole range of peak $E_{THz}$ in our experiments. As apparent
from this plot, the time delay between the reference and the
sample THz pulses in \emph{n}-GaAs increases with increasing peak
$E_{THz}$, which is a direct signature of SPM. The dashed black
line in Fig. \ref{fig2}(b) shows the estimated group delay
(i.e."delay as a whole") \cite{dmt_mch_THz_satabs} of a THz pulse in the material as a function of peak $E_{THz}$. In Figs. \ref{fig2}(d,
e) the normalized reference and sample pulses for the two extreme
values of peak $E_{THz}$ of 9 kV/cm and 292 kV/cm are shown. The
difference between the two reference waveforms in Fig. \ref{fig2}
(d) is fairly small: there is no observable time shift, and there is only a very small change in the pulse shapes, which is due to non-ideality of the THz wire-grid polarizers used for adjustment of the peak THz field. The difference between the two sample waveforms in Fig. \ref{fig2} (e) is, however, quite noticeable. The sample waveform corresponding to a stronger $E_{THz}$ clearly experiences a larger time delay, as well as a certain reshaping. Fig. \ref{fig2} is therefore the direct time-domain observation of THz SPM occurring in a single-cycle regime. In Fig. \ref{fig2} (f) we show the connection between the saturable absorption and SPM in our sample, by plotting both the THz frequency-integrated amplitude transmission coefficient
 and an estimated group delay of the sample pulse as a function of peak $E_{THz}$. The observed increase in the transmission coefficient is more than two-fold, and the increase in the group delay exceeds 150 fs, making it a noticeable fraction of the duration of the THz cycle.

\begin{figure} [t]
\centering
   \includegraphics[width=6.5cm]{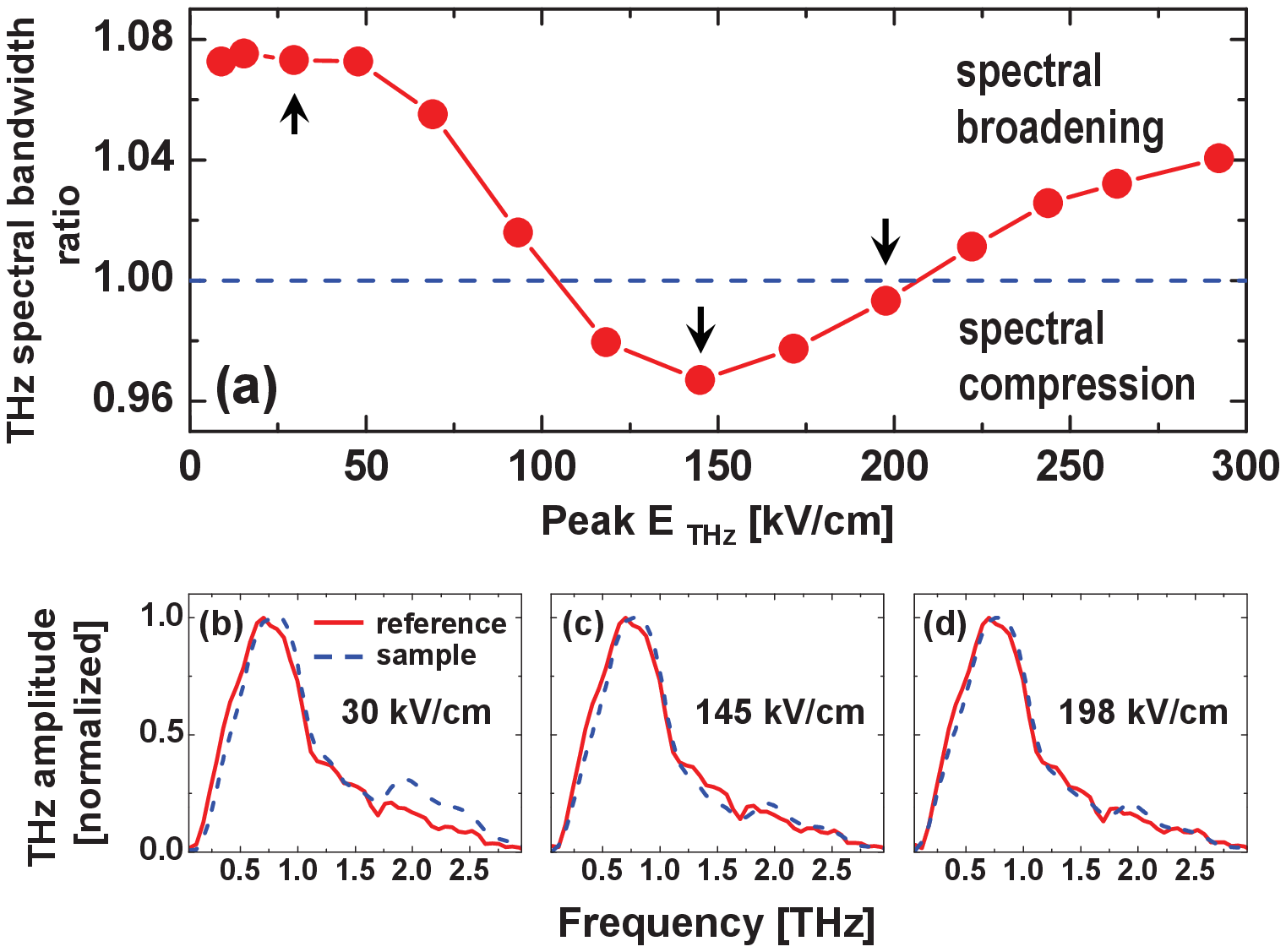}
  \caption
  {\label{fig4} Nonlinear spectral broadening and compression of a THz pulse in n-GaAs. (a) The ratio of effective bandwidths of the
  sample and reference pulses, as a function of peak $E_{THz}$. (b-d) Examples of spectral dynamics: normalized amplitude spectra of reference and sample THz pulses measured at peak THz field values of of 30 kV/cm, 145 kV/cm, and 198 kV/cm, indicated by arrows in
  (a). }
   \end{figure}

We now transfer to the frequency domain by using the Fourier
transforms of the measured THz pulses from Fig. \ref{fig2}
\cite{tonouchi, jepsen_LPR}. In Fig. \ref{fig3} we show the SPM as
a nonlinear, THz-field dependent contribution to the
refractive index of \emph{n}-GaAs, calculated as a correction to the refractive index spectrum measured at the weakest peak $E_{THz}$ ,
$\Delta n(\omega) = n(\omega, E_{THz}) -  n(\omega, \text{9 kV/cm})$. A very strong, and also frequency-dependent, dynamics of $\Delta
n(\omega)$ is observed with increase in peak $E_{THz}$ from 9
kV/cm to 292 kV/cm. Interestingly, the refractive index
nonlinearity is found to be both \emph{negative} and
\emph{positive} within the bandwidth of the same
single-cycle THz pulse, making it quite a unique situation in
nonlinear optics (though quite understandable given the
ultra-broadband nature of single-cycle pulses). These negative and
positive trends in the dynamics of $\Delta n$ at the frequencies
below and above approximately 0.5 THz, with increase in peak $E_{THz}$, are shown with blue arrows in Fig. \ref{fig3}. The maximum refractive index changes in our measurements were as large as
$\Delta n$ = -0.13 $-$ +0.08, in the spectral range 0.3 - 2.4 THz.
This corresponds to the nonlinear phase delay $\Delta \tau_{\phi}
= \Delta n d / c$ in the range of -173 $-$ +107 fs acquired by the
THz signal at different frequencies in our 0.4 - mm thick sample, also shown in Fig. \ref{fig3}.

\begin{figure} [t]
\centering
   \includegraphics[width=6.5cm]{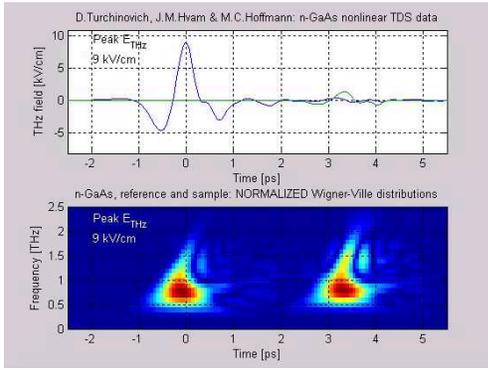}
  \caption
  {\label{movie_frame} A frame from the movie \cite{movie} showing the evolution of the reference and sample THz pulses in time domain, and their normalized Wigner-Ville spectrograms, with increase in peak $E_{THz}$.}
   \end{figure}

In our specific case most of the spectrum of the THz pulse (shown
as grey area in Fig. \ref{fig3}) was in the spectral range
of positive index nonlinearity, which results in the increasing group delay of the THz pulse at higher peak $E_{THz}$ (see Fig. \ref{fig2}). However, if the THz pulse spectrum belonged to the range below 0.5 THz, the decrease in group delay with increase in $E_{THz}$ would occur instead.

\begin{figure} [t]
\centering
   \includegraphics[width=6.5cm]{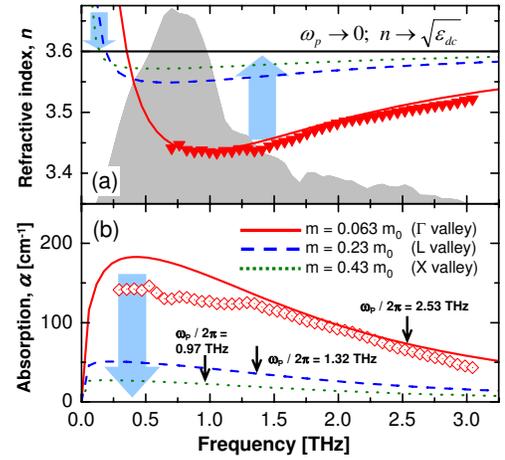}
  \caption
  {\label{fig5} Blue arrows: trends in (a) refractive index and (b) absorption THz spectra of \emph{n}-GaAs with decrease in $\omega_p$. Spectra measured in linear regime (symbols), and calculated using a DS model with  $N = 5 \times 10^{15}$ cm$^{-3}$, $\tau_{r}$ = 94 fs, and $c_{DS} = - 0.15$ for the electron effective mass of $\Gamma$- (red solid line), $L$- (blue dashed line), and $X-$ (green dotted line) valleys. Amplitude spectrum of a reference THz pulse is shown in the background of (a). Black arrows indicate the corresponding $\omega_p$ values. Horizontal line in (a) corresponds to $\omega_p = 0$, when $n = \sqrt{\varepsilon_{dc}}$ = 3.6.}
\end{figure}

Inherent to SPM is spectral reshaping of the optical signal
\cite{Agrawal, Hasnain}, which can lead to both spectral
broadening \emph{and} compression within the same
NLO system (see e.g. \cite{dmt_OE_PCF_laser}). In
Fig. \ref{fig4}(a) it is illustrated by the dependency of the ratio of effective spectral bandwidths of the sample and reference pulses $\Delta \omega_{sam} / \Delta \omega _{ref}$ on peak $E_{THz}$ \cite{effective_bandwidth}. If this ratio is greater (smaller) than unity, then relative spectral broadening (compression) of a THz pulse in a sample takes place. We observed a clear spectral breathing by approx. -3 $-$ +7 \% of a THz pulse in our sample. In Figs. \ref{fig4}(b-d) the normalized amplitude spectra of reference and sample THz pulses at selected values of peak $E_{THz}$ are shown, exemplifying the spectral broadening, compression, and virtually no change, respectively, of THz pulses in \emph{n}-GaAs. In general, the observed spectral breathing will depend on the combination of pulse chirp, index nonlinearity, and frequency-dependent nonlinear absorption, which can only be quantified using advanced numerical methods \cite{Koch, koch_hegmann}.

A cognitive correlation between the THz pulse transmission and delay as a function of peak $E_{THz}$, also depicting the evolution of THz pulse spectrograms \cite{Zhang_spectrograms}, is visualized by a \emph{movie}, which can be found under the Ref. \cite{movie}. An example of the movie frame is shown in Fig. \ref{movie_frame}.

We note that the link between the refractive index and absorption (or gain) nonlinearities is fundamental, and is used e.g.  in semiconductor slow/fast light devices for the telecom range (see e.g. \cite{mork, Hasnain}). We also note that the reshaping of a THz signal generated in a LN crystal under variable optical pump strength was reported and assigned to THz SPM in \cite{hebling_SPM}, but later studies clearly related this effect to the optical-range nonlinearity of the laser pump pulse in LN \cite{tanaka_chi2chi2}. The reshaping by SPM of a THz pulse in InSb undergoing impact ionization was reported \cite{Lindenberg}, however, without the systematic analysis.

Now we illustrate the observed trends in nonlinear absorption and refractive index by simple modeling. In Fig. \ref{fig5} the measured \emph{linear} THz spectra of our sample are shown (symbols) along with the Drude fit using the parameters $N$ and $\tau_r$ described above,  $n = \sqrt{\varepsilon_{dc}}$ = 3.6 \cite{Grischkowsky}, and $m_e = m_{\Gamma}$ (solid lines). For this fit we used a Drude-Smith (DS) model incorporating the effective carrier localization into an ideal Drude model \cite{jepsen_LPR, Smith, Nemec}. The small localization parameter $c_{DS} = - 0.15$ used here shows that our sample is very close to an ideal Drude plasma system described by Eq. \ref{eq1}. In the same figure we show the calculated $n$ and $\alpha$ spectra for the same sample parameters as before, but using the heavier effective masses $m_e = m_{L}$ and $m_e = m_{X}$ of the satellite valleys \cite{ponderomotive} (dashed and dotted lines, respectively). Corresponding values for $\omega_p$ are shown in the figure using black arrows. It is clear that with decrease in $\omega_p$ the absorption decreases monotonously. The trend in refractive index with decreasing $\omega_p$ is the decrease from larger values at lower frequencies ($<$ 0.5 THz), and increase from lower values at higher frequencies, approaching the background value of $n = \sqrt{\varepsilon_{dc}}$ at $\omega_p = 0$. These trends in the dynamics of $\alpha$ and $n$, indicated by the blue arrows, exactly match the observed saturable absorption (Fig. \ref{fig2}(f)), and the behavior of refractive index nonlinearity (Fig. \ref{fig3}), including its opposite signs at lower and higher frequencies. The large value of refractive index at lower frequencies in a metal or doped semiconductor directly follows from Eq. \ref{eq1}: it is known as Hagen-Rubens regime, and is characteristic for the frequency range  $\omega \ll \tau_{r}^{-1}$ \cite{Dressel}. The spectral position of zero index nonlinearity is thus defined by the electron momentum relaxation rate $\tau_{r}^{-1}$, and can be theoretically predicted with reasonable accuracy, as shown by Figs. \ref{fig3} and \ref{fig5}(a).

In conclusion, we have studied the SPM of a single-cycle THz pulse in a semiconductor, whose THz-range optical nonlinearity is based on the carrier heating in THz fields, and is adequately described by the Drude plasma model.

We are grateful to K. Yvind, J. L{\ae}gsgaard, J. M{\o}rk (DTU
Fotonik) and A. Cavalleri (Univ. Hamburg) for valuable
discussions. We acknowledge the partial financial support from the
Danish Proof of Concept Foundation (grant 7.7 Ultra-high-speed
wireless data communications), Danish Council for Independent
Research - Technology and Production Sciences (FTP), and Max
Planck Society.


\begin{thebibliography}{99}
\bibitem{justification_epsilon_dc} Here we can ignore, for simplicity, the dispersion due to a phonon contribution to $\hat{\varepsilon}$, which is generally independent of free carriers.
\bibitem{ponderomotive} A static Gunn field of GaAs is about 5 kV/cm. A THz signal with the central frequency of 1 THz and with the peak $E_{THz}$ = 100 kV/cm has a ponderomotive energy of about 1.8 eV, clearly exceeding the $\Gamma$-$L$ and $\Gamma$-$X$ energy separations of 0.33 eV and 0.52 eV, respectively \cite{Brennan_Hess}, thus making intervalley transitions possible.
\bibitem{effective_bandwidth} $\Delta \omega = \int_{\omega} E(\omega) d{\omega} /  \text{max}\{E(\omega)\}$, where $E(\omega)$ is Fourier amplitude of the corresponding THz pulse.
\end{thebibliography}
\end{document}